\documentclass{PoS}

\usepackage{epsfig,wrapfig}

\title{QCD and Top Quark Physics at the LHC}
\ShortTitle{QCD and Top Quark Physics at the LHC}

\author{
Frank-Peter SCHILLING
\thanks{On behalf of the ATLAS and CMS Collaborations}\\
Institut f\"ur Experimentelle Kernphysik, University of Karlsruhe / KIT, Germany\\
E-mail: \email{frank-peter.schilling@cern.ch}}

\abstract{
The expected performance of the ATLAS and CMS detectors at the Large
Hadron Collider (LHC) in QCD and top quark measurements is discussed,
with a focus on the early data taking phase. Such processes are
amongst the primary backgrounds in the searches for new physics, and
thus must be understood very well before discoveries can be made.  In
addition, they serve as useful detector calibration candles.  }

\FullConference{Physics at LHC 2008\\
                29 September - 04 October, 2008\\
                Split, Croatia}

\begin{document}


\section{Introduction}

The prime goal of the ATLAS and CMS experiments at the Large Hadron
Collider LHC at CERN is to find the Standard Model Higgs boson and to
look for possible signatures of new physics beyond the Standard Model,
such as Supersymmetry. However, before discoveries can be made, the
detectors have to be calibrated, and the Standard Model backgrounds
have to be understood. The LHC will produce QCD events with jets at a very
high rate and with a wide kinematic range. In addition, the production
cross section for top quark pairs is enhanced by two orders of
magnitude compared with the Fermilab TEVATRON.

QCD jet and top quark processes will be used for Standard Model
measurements and tuning og Monte Carlo (MC) generators.  In addition
di-jet and top quark final states are important tools for calibrating
the detectors.  In this article, the potential of ATLAS and CMS in the
fields of QCD and Top Quark Physics is briefly summarized, with the
focus on recently performed studies using the most recent simulation
and reconstruction software of the experiments, and which are targeted
at the first LHC data.


\section{QCD Physics}

QCD processes such as multi-jet production will occur at a very high
rate, and in a kinematic range (high jet transverse momentum $p_T$,
low Bjorken-$x$) which is considerably extended with respect to
e.g. the Fermilab TEVATRON. The data can be used to measure jet cross sections
and to place constraints on the parton density functions (PDFs) of the
proton and on the strong coupling $\alpha_s$. At low $p_T$, Monte
Carlo models for multiple interactions and the underlying event can be
tuned, whereas at high $p_T$, deviations from the expected behavior
may be due to new physics signals.

\subsection{Low $\mathbf{p_T}$ QCD}

\begin{figure}
\centering
\begin{minipage}{0.4\linewidth}
\epsfig{file=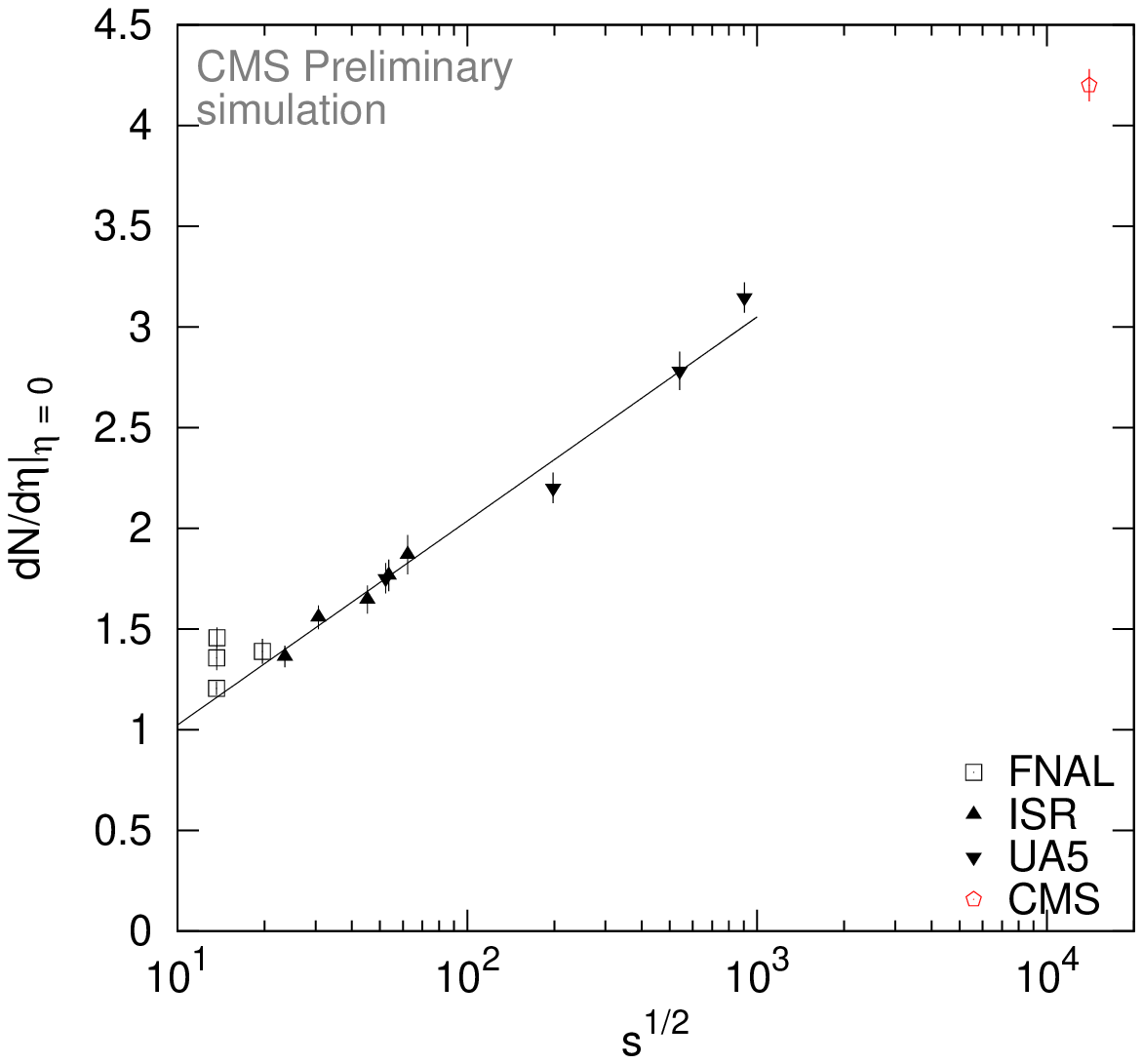,width=0.99\linewidth}
\end{minipage}
\begin{minipage}{0.4\linewidth}
\vspace{-2.0mm}
\epsfig{file=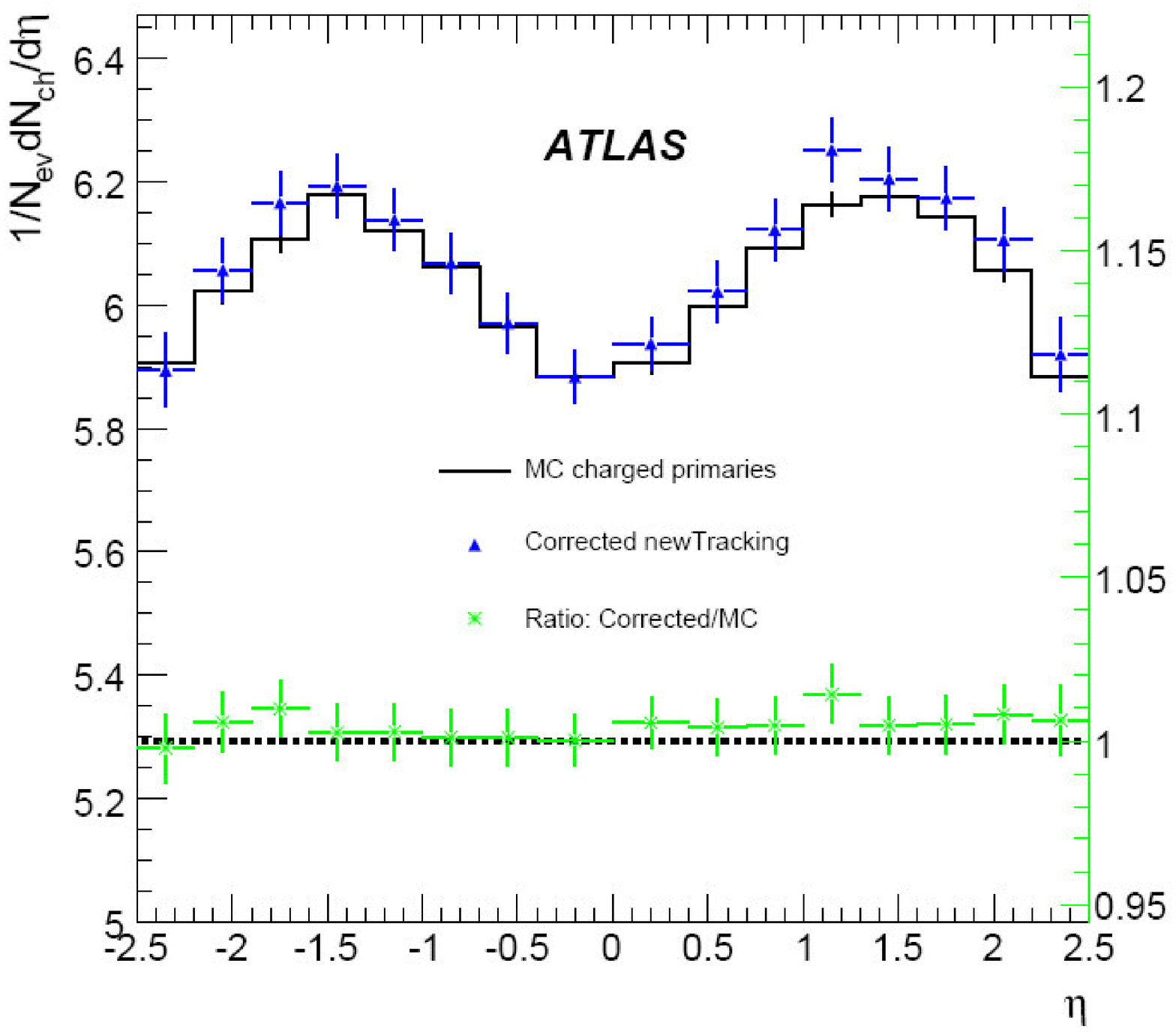,width=0.99\linewidth}
\end{minipage}
\caption{(left) Energy dependence of pseudo-rapidity density
of charged hadrons at $\eta\sim0$ (CMS). 
(right) Corrected normalized charged particle pseudo-rapidity distribution
for non-single-diffractive events (ATLAS).
}
\label{fig:lowpt}
\end{figure}

Cross sections and differential yields of charged particles produced
at LHC energies can be measured with good precision with the ATLAS and
CMS tracking detectors. In a recent CMS study~\cite{mult-cms}, around
2 million simulated minimum bias events are used, corresponding to 1
Hz of trigger rate collected over a period of one month of
data-taking.  Tracks are reconstructed with transverse momenta as low
as $p_T = 75 \rm\ MeV$ using a dedicated pixel hit triplet seeding
technique. The measured distributions are corrected for acceptance, trigger
efficiency etc., and the systematic error is around
$8\%$. Figure~\ref{fig:lowpt} (left) shows the projected measurement
of the charged particle density at $\eta\sim0$, which would allow to
discriminate between models predicting either a $\sim \ln(s)$ or $\sim
\ln^2(s)$ behavior. Figure~\ref{fig:lowpt} (right) shows a similar
study by ATLAS~\cite{atlas-csc} of a measurement of the charged
particle density vs. $\eta$ for $p_T>150 \rm\ MeV$ for
non-single-diffractive events.

\begin{figure}
\centering
\begin{minipage}{0.45\linewidth}
\vspace{1.0mm}
\epsfig{file=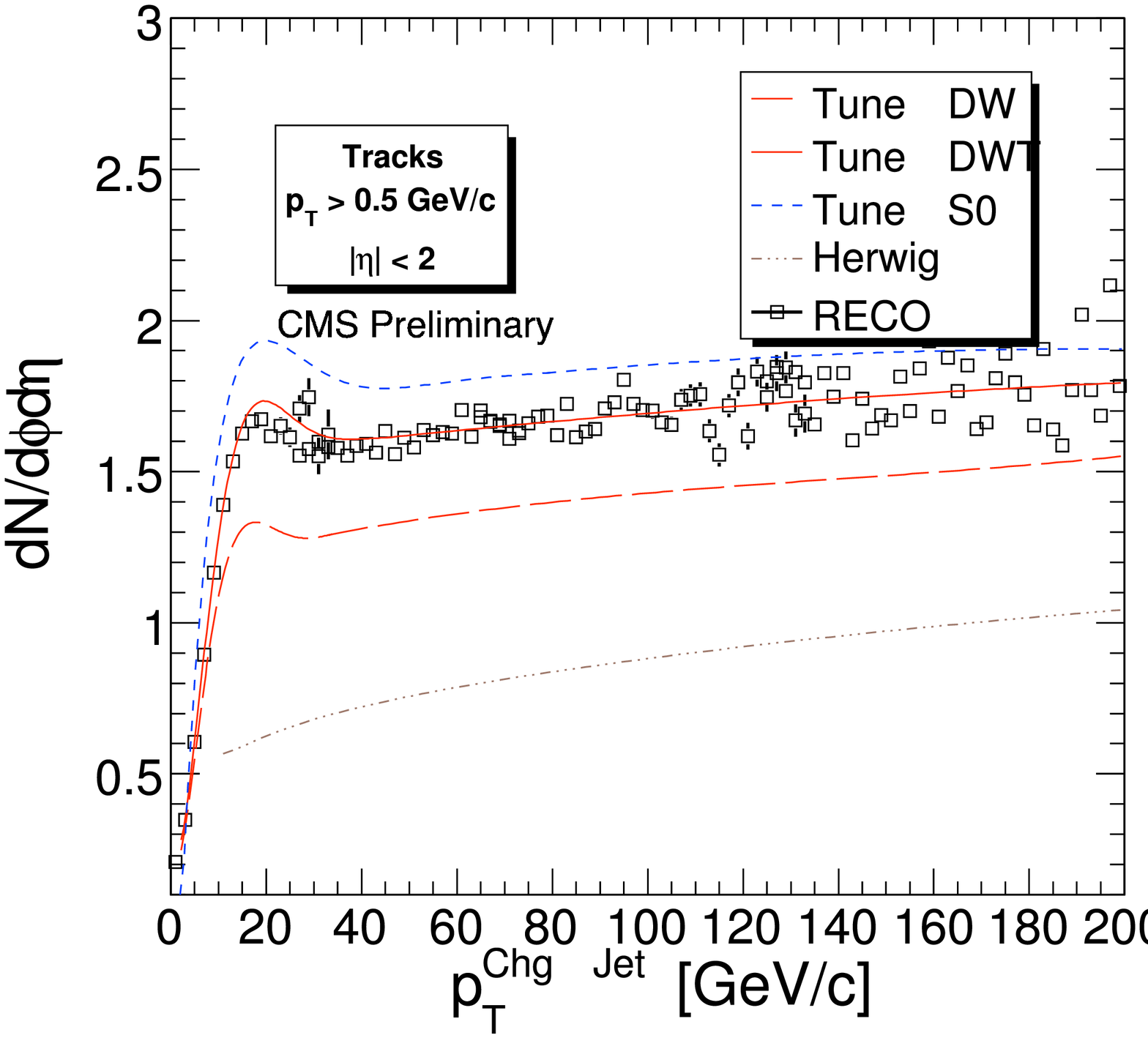,width=0.99\linewidth,height=4.7cm}
\end{minipage}
\begin{minipage}{0.45\linewidth}
\epsfig{file=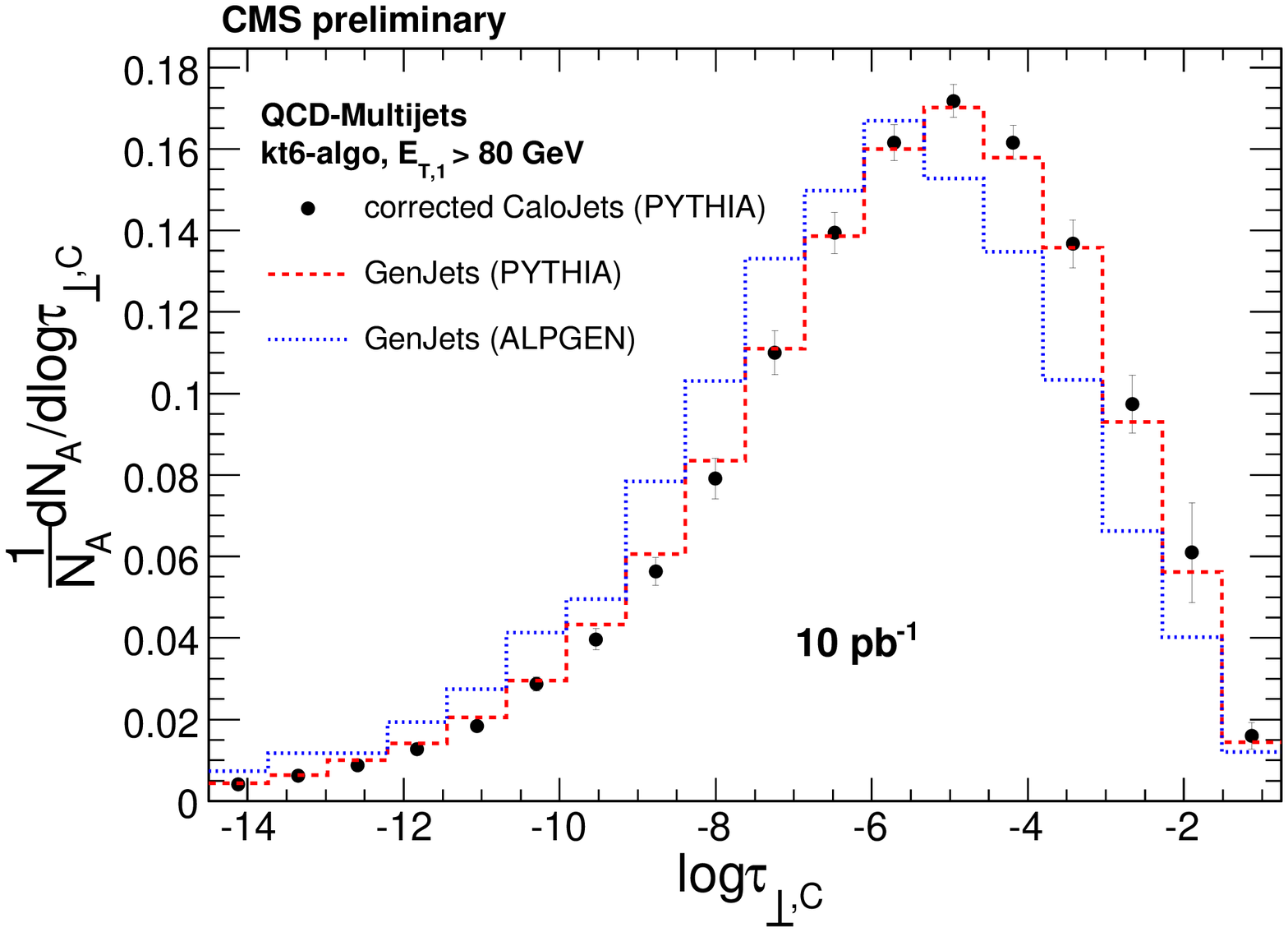,width=0.99\linewidth}
\end{minipage}
\caption{
(left) Corrected charged particle density in the transverse region as a function of the leading track jet $p_T$ (CMS).
(right) Central transverse thrust distribution for $10\rm\ pb^{-1}$
(CMS).
}
\label{fig:lowpt2}
\end{figure}

Measurements of the underlying event (defined as all activity but the
hard scatter) at LHC energies are crucial for the understanding of
multi parton interactions as well as for the tuning of Monte Carlo
models.  It can be studied by looking at the hadronic activity in the
region transverse to the leading jet in the event. Extrapolations of
existing models predict an activity which is two times higher than that at
the TEVATRON. Figure~\ref{fig:lowpt2} (left) shows the result of a CMS
study~\cite{mbue-cms} for $10 \rm\ pb^{-1}$. The plot shows the
corrected charged particle density in the transverse region as a
function of the leading track jet $p_T$, where a high sensitivity with
respect to the various available MC tunes can be observed.

\subsection{High $\mathbf{p_T}$ QCD}

Both collaborations extensively study QCD jet production using the
most advanced jet algorithms such as SIS-Cone~\cite{siscone} and
Fast-Kt~\cite{fastjet}, using either calorimeter towers, tracks or
particle flow objects as input to the jet finding. For the jet energy
scale (JES) calibration, CMS uses a factorized multi-level jet energy
correction approach~\cite{jetreco-cms}, whereas ATLAS calibrates
calorimeter clusters to the particle scale, which are then used as
input for the jet finding. Both experiments optionally apply
corrections to the parton level.

Dijet azimuthal decorrelations have been studied by
ATLAS~\cite{atlas-dphi}, by looking at the difference in $\phi$
between the leading two jets in dijet events.  This quantity is very
sensitive to gluon radiation and higher order effects, and allows
tuning of MC models.

Normalized event shapes, using collinear and infrared safe
observables, are robust against JES uncertainties and are thus
suitable for the early data analysis. Figure~\ref{fig:lowpt2} (right)
shows a projection for $10\rm\ pb^{-1}$ of CMS
data~\cite{evshapes-cms} of the corrected central transverse thrust
distribution for jets with $p_T>80 \rm\ GeV$.  The simulated data are
compared with two predictions using either PYTHIA or ALPGEN.

\begin{wrapfigure}{r}{0.5\linewidth}
\epsfig{file=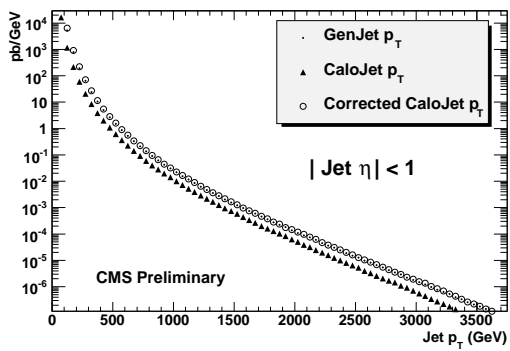,width=\linewidth}
\caption{Inclusive jet cross section vs $p_T$ for $10\rm\ pb^{-1}$ 
simulated data (CMS).
}
\label{fig:highpt}
\end{wrapfigure}

The measurement of the inclusive jet cross section is one of the most
important early measurements at the LHC. Due to the high cross
section, it can be measured up to very high $p_T$ values already at
the beginning of data taking. Figure~\ref{fig:highpt} shows a
prediction for $10\rm\ pb^{-1}$ of CMS data~\cite{jetxs-cms}, using
jets with $|\eta|<1$. Such measurements will allow to test
perturbative QCD calculations in a new regime, place constraints on PDFs and
$\alpha_s$, and probe signatures of new physics (e.g. quark compositeness).

ATLAS recently performed a detailed
investigation~\cite{jeterrors-atlas} of the systematic errors
associated with the inclusive jet cross section measurement. The
dominating error is clearly due to the knowledge of the jet energy
scale: An error of 5 (10) \% on the jet energy scale leads to a cross
section uncertainty of 35 (70) \% at 1 TeV and 50 (100) \% at 2 TeV.
On the other hand, the statistical uncertainty (for a bin width of 100
GeV) is much smaller, namely 2 (30) \% at 1 (2) TeV for an integrated
luminosity of $100 \rm\ pb^{-1}$.  The theoretical uncertainty
originating from the scale uncertainty of the next-to-leading order
(NLO) calculation, obtained by varying the factorization and
renormalization scales $\mu_f=\mu_r$ between $p_T/2$ and $2p_T$, is
estimated as $5-10\%$ at 1 TeV.


\section{Top Quark Physics}

The LHC will be a top quark factory: at 14 TeV centre-of-mass energy,
around 8M $t\bar{t}$ pairs will be produced per year of data taking
($10 \rm\ fb^{-1}$). At LHC the $t\bar{t}$ production is predominantly
gluon induced, in contrast with the TEVATRON where top is mostly
produced from (anti-) quarks in the proton, and the cross sections, both for
pair production as well as for electroweak single production of top
quarks, will be increased by about two orders of magnitude.

The LHC top quark physics program consists first of the rediscovery of
the top quark, followed by (differential) cross section measurements,
measurements of the properties of the heaviest Standard Model fermion
(mass, spin etc.), as well as the search for new heavy resonances
decaying into $t\bar{t}$ pairs. Besides the fundamental interest in
the top quark and its properties itself, top production constitutes
one of the major backgrounds in many scenarios of new physics, such as
supersymmetry. Moreover, top events are very useful for the
understanding of the detector, as they involve most of the physics
objects (leptons, jets, missing $E_T$ (MET), b-tagging), and can also be
used to constrain the JES and the b-tagging efficiency.

\begin{figure}
\centering
\begin{minipage}{0.44\linewidth}
\epsfig{file=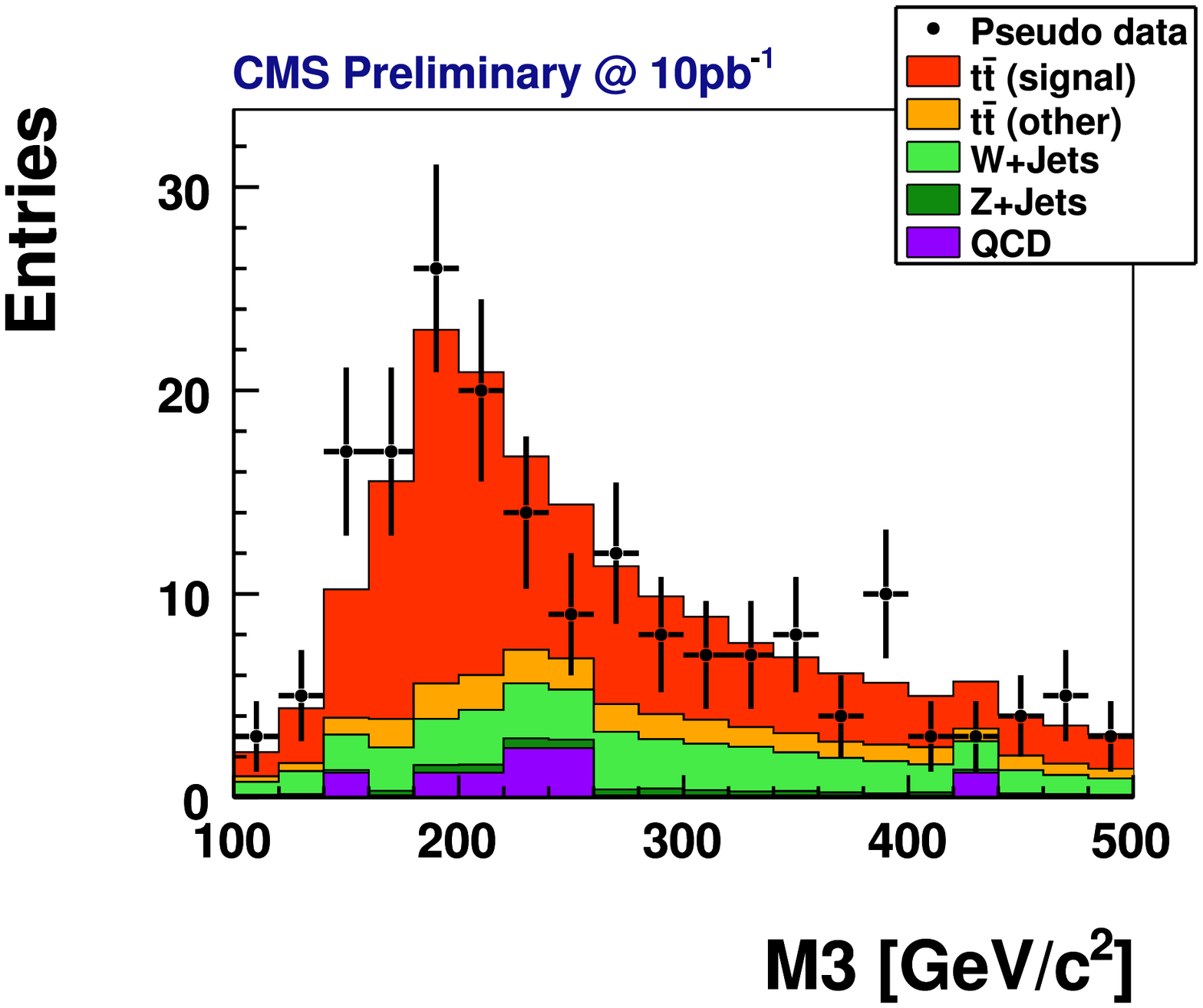,width=0.8\linewidth,angle=90}
\end{minipage}
\begin{minipage}{0.55\linewidth}
\vspace{-1.0mm}
\epsfig{file=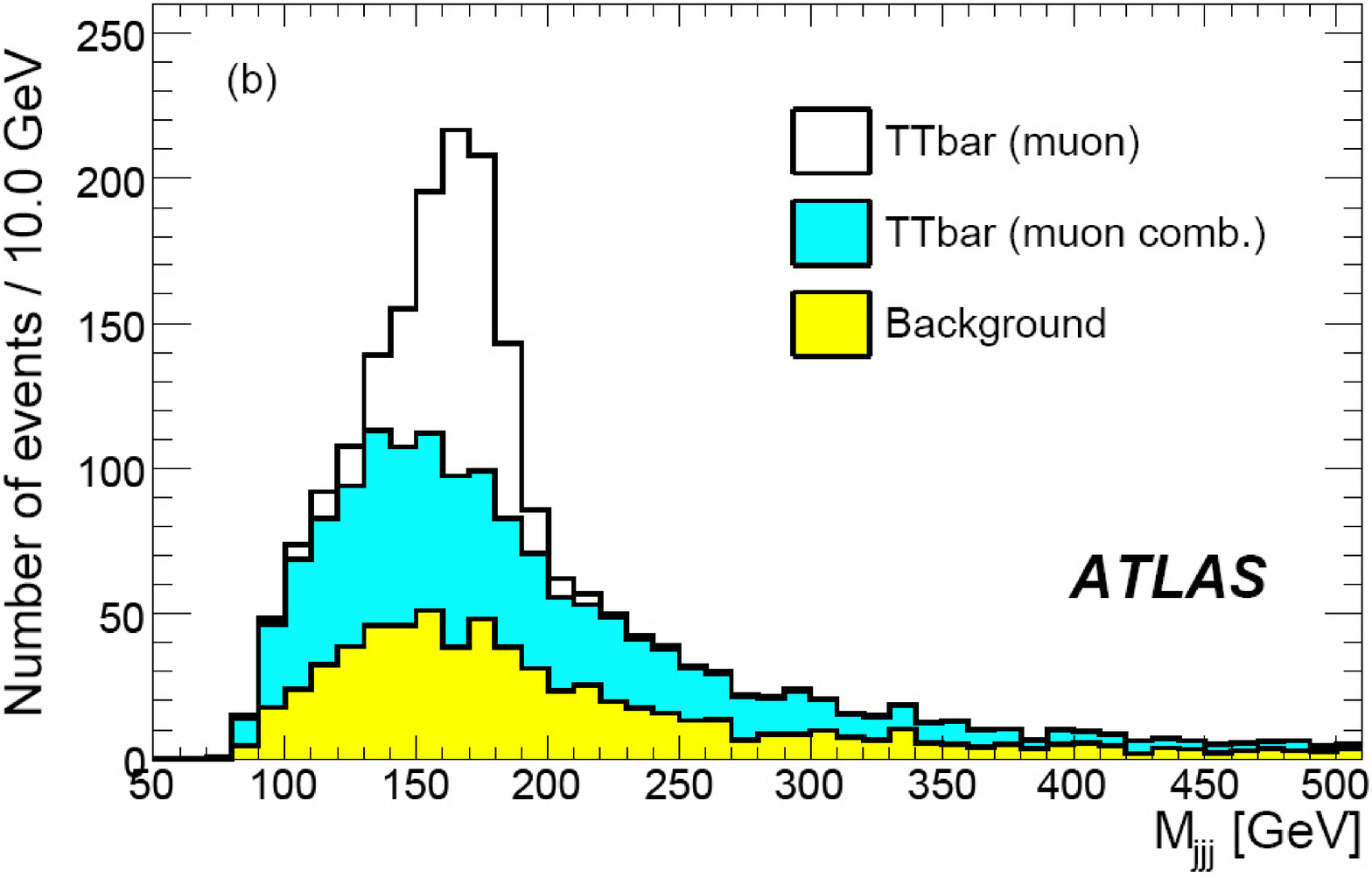,width=0.8\linewidth}
\end{minipage}
\caption{(left) Invariant mass of the three jets with the highest
vectorially summed $E_T$ in the semileptonic muon channel 
for $10 \rm\ pb^{-1}$ (CMS). (right)
The same quantity for $100 \rm\ pb^{-1}$ (ATLAS).}
\label{fig:top1}
\end{figure}

\subsection{$\mathbf{t\bar{t}}$ Cross Section}

The $t\bar{t}$ production cross section has been
estimated~\cite{topxs} as $\sigma(t\bar{t}) = 908\pm 83 {\rm (scale)}
\pm 30 {\rm (PDF)} \rm\ pb$ at 14 TeV and $\sigma(t\bar{t}) = 414\pm
40 {\rm (scale)} \pm 20 {\rm (PDF)} \rm\ pb$ at 10 TeV (using NLO
calculations with next-to-leading-log (NLL) resummation, $m_t=171 \rm\ GeV$ and the CTEQ 6.5
PDFs).
CMS has studied the rediscovery of the top quark in the semileptonic
muon channel~\cite{topslmuon-cms} with a simple selection without
b-tagging or MET requirement. 
Requesting one isolated muon with $p_T>30 \rm\ GeV$
and at least 4 jets with $p_T>65,40,40,40 \rm\ GeV$, around 128
$t\bar{t}$ signal and 88 background (mostly W+jets) events are
expected for an integrated luminosity of $10 \rm\ pb^{-1}$, see
Figure~\ref{fig:top1} (left). ATLAS has investigated~\cite{atlas-csc}
the lepton+jets channel (electron or muon) for $100\rm\ pb^{-1}$,
requesting one lepton with $p_T>20 \rm\ GeV$, at least 4 jets with
$p_T>40,40,40,20 \rm\ GeV$, ${\rm MET}>20 \rm\ GeV$ and
b-tagging. They expect 755 (143) signal (background) events without
b-tag and 403 (42) signal (background) events when requiring at least
one b-tag jet.
The cross section is extracted from a fit to the invariant mass
distribution, see Figure~\ref{fig:top1} (right), and the expected
significance is around 9 for $100\rm\ pb^{-1}$.

\begin{figure}
\centering
\begin{minipage}{0.4\linewidth}
\epsfig{file=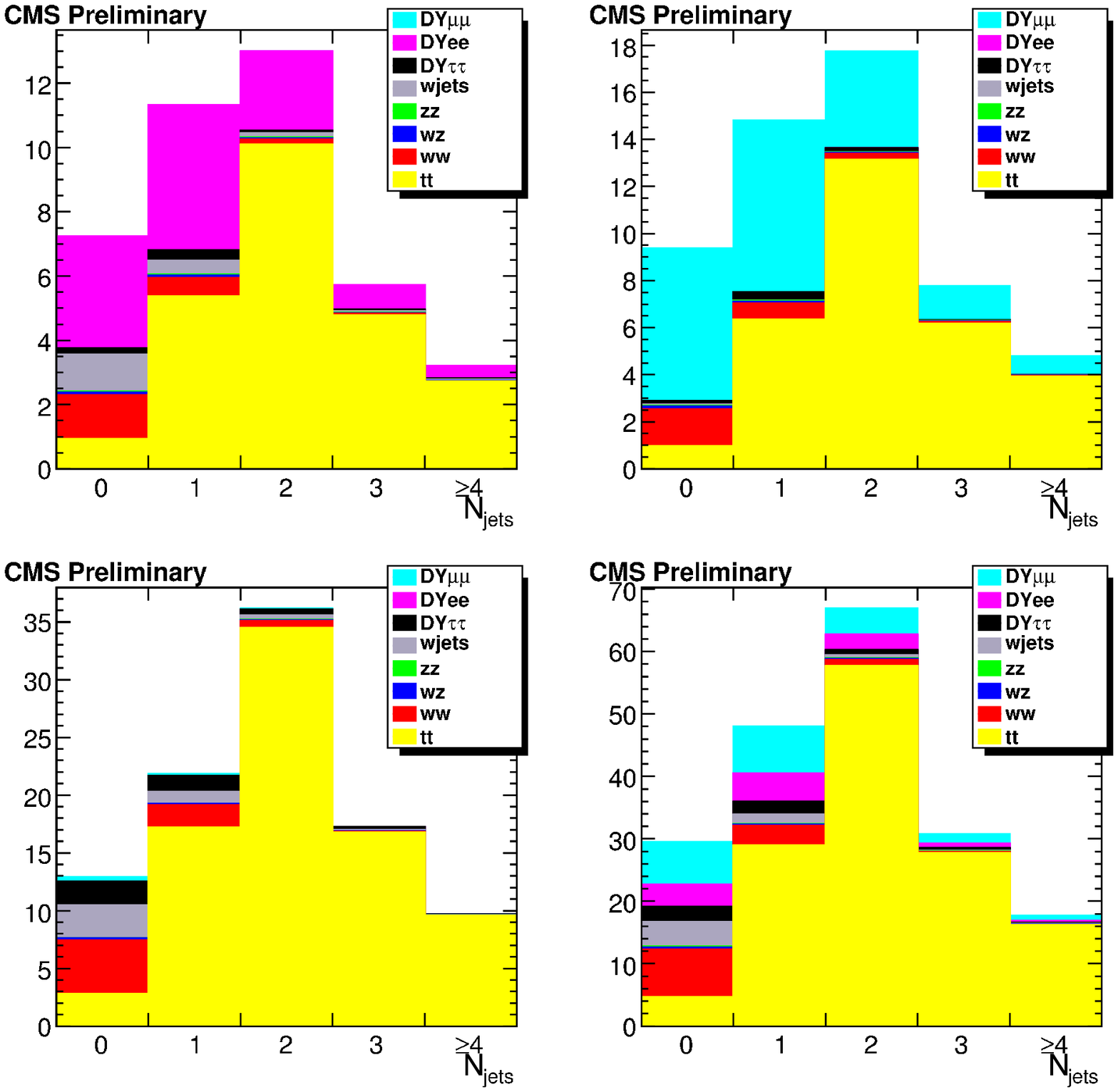,width=0.8\linewidth,clip=}
\end{minipage}
\begin{minipage}{0.5\linewidth}
\epsfig{file=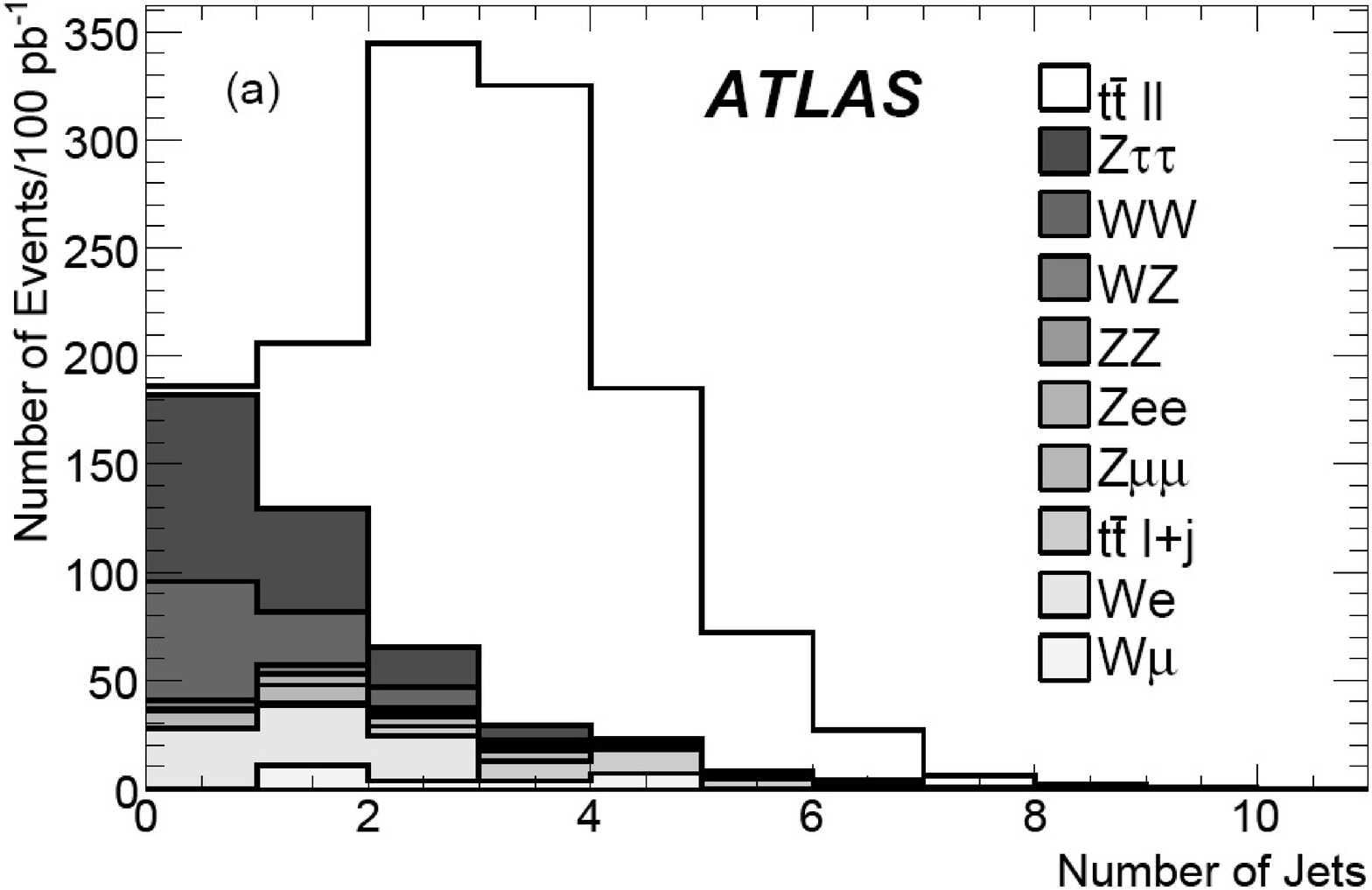,width=0.9\linewidth}
\end{minipage}
\caption{
Jet multiplicity in the top dilepton channel for 
$10 \rm\ pb^{-1}$ of simulated CMS data (left) and for
 $100 \rm\ pb^{-1}$ of simulated ATLAS data (right). 
}
\label{fig:top2}
\end{figure}

In the dilepton channel, where QCD background is not an issue, CMS
expects~\cite{topdilep-cms} a statistical uncertainty on the number of
$t\bar{t}$ events of around 10\% for $L=10\rm\ pb^{-1}$, combining
$ee$, $\mu e$ and $\mu\mu$ channels, see Figure~\ref{fig:top2} (left).
Events are selected requesting two leptons with $p_T>20 \rm\ GeV$, at
least two jets with $p_T>30 \rm\ GeV$, a MET cut and a Drell-Yan
veto. With softer jet cuts ($p_T>20 \rm\ GeV$), ATLAS
estimates~\cite{atlas-csc} that the cross section can be measured in
this channel with a precision of $4 {\rm (stat.)} \pm 4 {\rm (syst.)}
\pm 2 {\rm (PDF)} \pm 5 {\rm (lumi)} \%$ for $100 \rm\ pb^{-1}$
(Figure~\ref{fig:top2} right).

The expected performance of the measurement of the top quark mass
is discussed in~\cite{topmassproc}.

\begin{figure}
\centering
\begin{minipage}{0.45\linewidth}
\epsfig{file=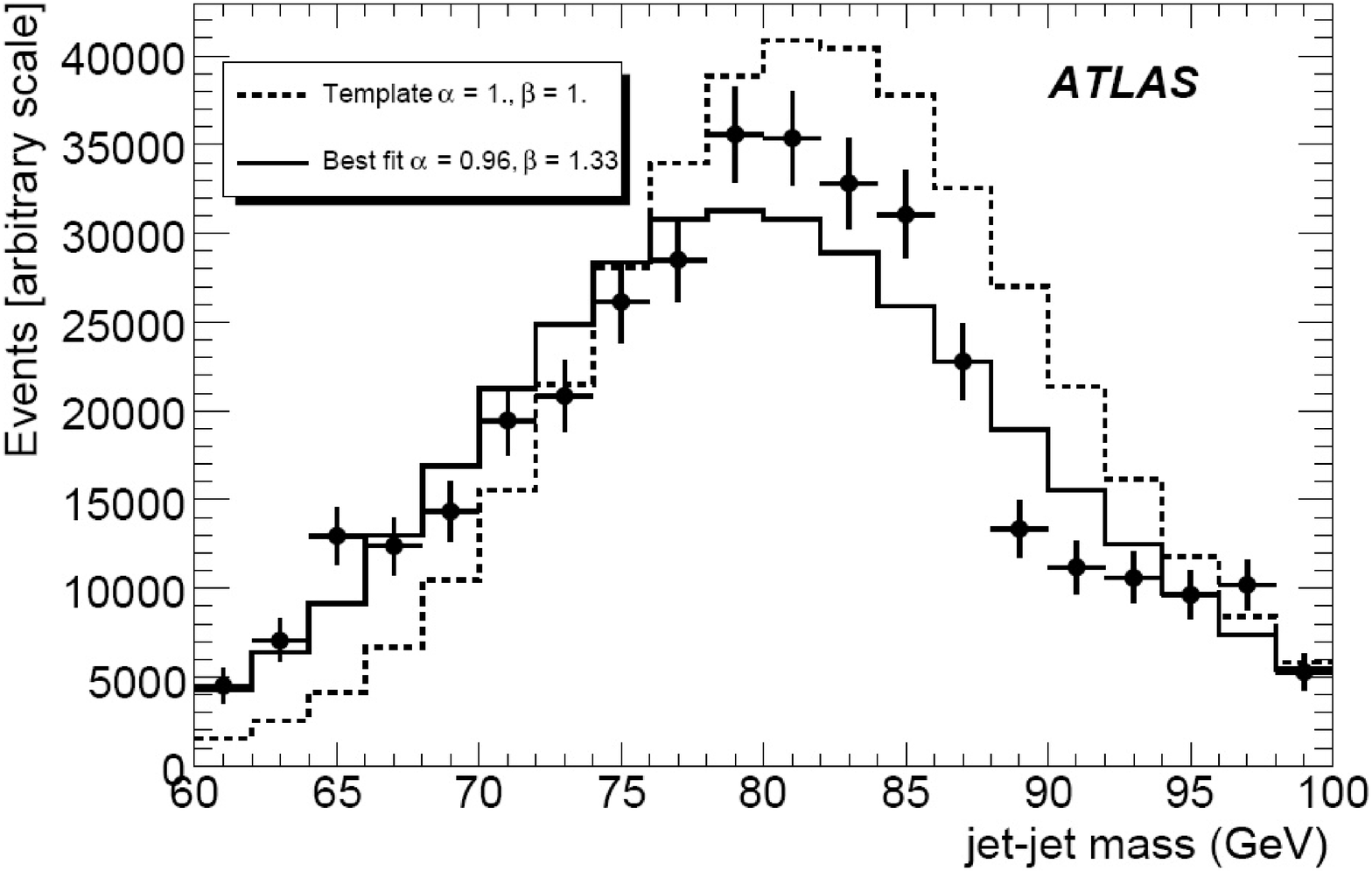,width=0.99\linewidth,clip=}
\end{minipage}
\begin{minipage}{0.45\linewidth}
\epsfig{file=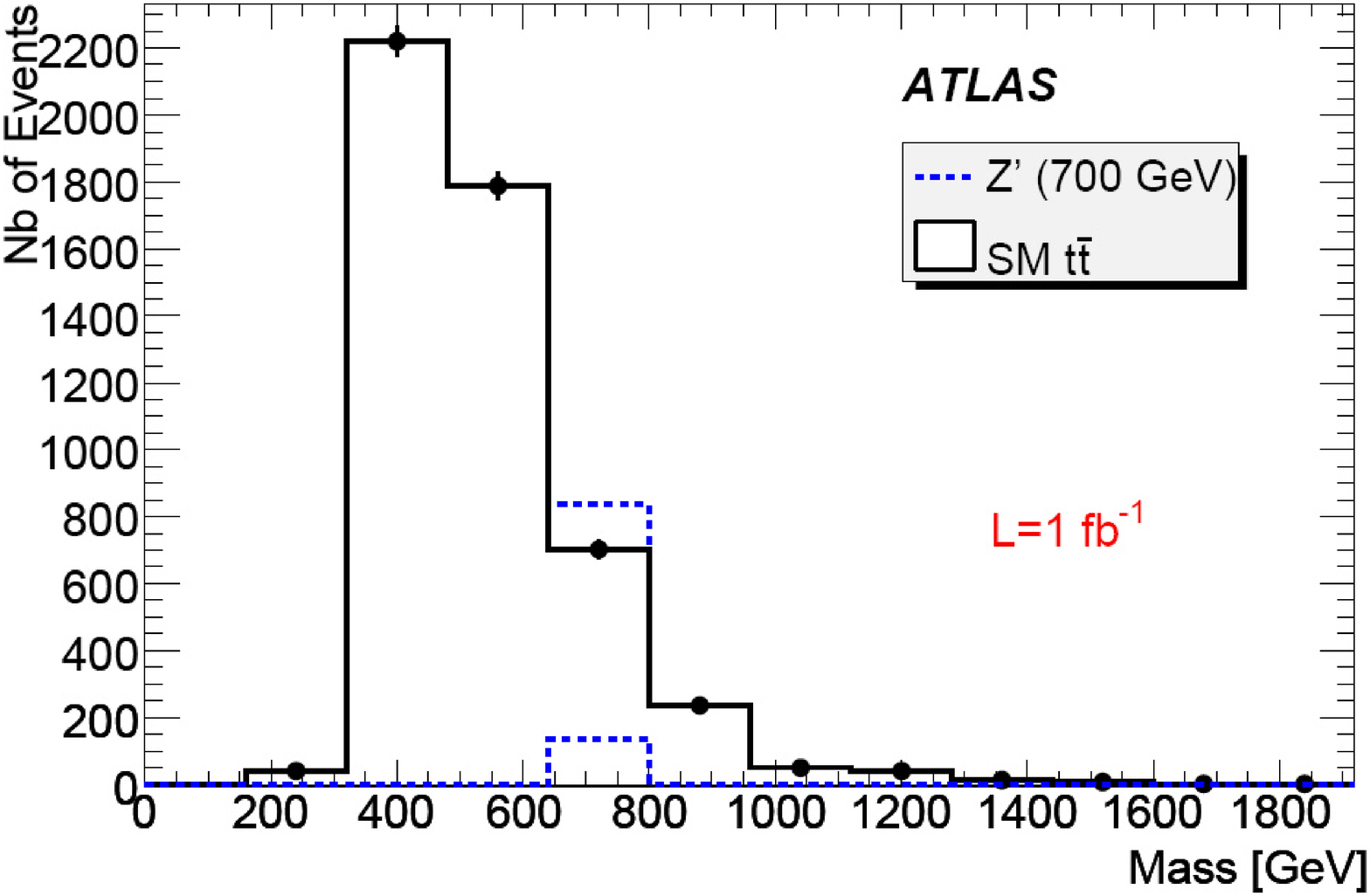,width=0.95\linewidth}
\end{minipage}
\caption{
(left) Dijet invariant mass in semileptonic $t\bar{t}$ events as used in 
a template fit to determine the light JES in ATLAS. 
(right) $m_{t\bar{t}}$ distribution showing a 
$Z'$ signal ($m_{Z'} = 700 \rm\ GeV$, $\sigma * {\rm BR} = 11 \rm\ pb$) 
on top of the Standard Model $t\bar{t}$ background (ATLAS $1 \rm\ fb^{-1}$).
}
\label{fig:top3}
\end{figure}

\subsection{Top as Calibration Tool}

$t\bar{t}$ events are also very useful calibration candles. ATLAS has
studied~\cite{atlas-csc} a determination of the light jet energy scale
using an $m_W$ constraint on the invariant mass of the two light jets
in semileptonic $t\bar{t}$ events. A precision of $2\%$ can be
achieved with $50\rm\ pb^{-1}$ of data (Figure~\ref{fig:top3} left).
A similar CMS study~\cite{topjes-cms} estimates an error of $1\%$ for
$100\rm\ pb^{-1}$ of data.  A high purity selection of $t\bar{t}$
events can also be used to determine the b-tagging efficiency with a
precision of $6\%$ in $1 \rm\ fb^{-1}$ of CMS data~\cite{topbtag-cms}.

\begin{wrapfigure}{r}{0.4\linewidth}
\epsfig{file=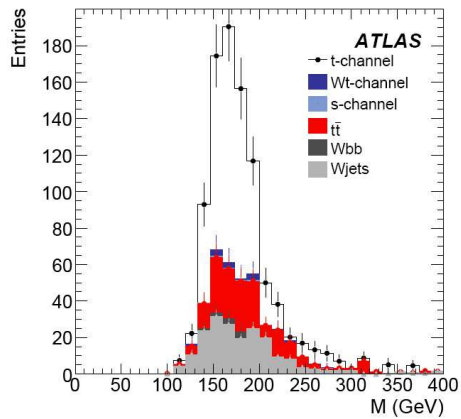,clip=,width=\linewidth}
\caption{Reconstructed top invariant mass in the single
top t-channel in the ATLAS study for $1 \rm\ fb^{-1}$.}
\label{fig:top4}
\end{wrapfigure}

\subsection{$\mathbf{t\bar{t}}$ Resonances} 

Due to the large Yukawa coupling, there is a significant potential to
discover new physics by searching for a new heavy resonances decaying
into a $t\bar{t}$ pair, which would manifest themselves as peaks or shape
distortions in the $m_{t\bar{t}}$ distribution.  There are several
experimental issues involved in reconstructing $m_{t\bar{t}}$. Due to
the boost the top quark decay products are close by, which means that
lepton isolation will deteriorate, and jets are merged. Furthermore the
b-tagging performance will decrease due to the high track density in the
jets.  These issues are currently being addressed by both collaborations.
Using a standard event reconstruction and selection in the lepton+jets
channel, ATLAS estimates to exclude Kaluza-Klein gluon resonances up
to $M\sim 1.5 \rm\ TeV$ with $1 \rm\ fb^{-1}$~\cite{atlas-csc}.  A $Z'$
signal with $m_{Z'} = 700 \rm\ GeV$ and $\sigma * {\rm BR} = 11 \rm\ pb$ is
shown in Figure~\ref{fig:top3} (right).

\subsection{Single Top Production} 

The single top (and antitop) production cross section at 14 TeV has
been estimated as 10 / 247 / 56 pb in the s / t / tW channels,
respectively~\cite{singletop-xs1,singletop-xs2}. It is thus around 1/3
of the size of the $t\bar{t}$ production. The main backgrounds are
W+jets and $t\bar{t}$, which need to be controlled using data driven
techniques. In the context of low signal over background ratio, the
use of sophisticated tools such as genetic algorithms, likelihoods and
Boosted Decision Trees will be very useful.  A precise determination of the
single top cross section can be achieved with a few $\rm fb^{-1}$ in
the t- (see Figure~\ref{fig:top4}) and Wt-channels.  Recent ATLAS
studies~\cite{atlas-csc} predict a signal to background ratio of 1.3
(0.3) in the t (Wt) channel using 1 (10) $\rm fb^{-1}$ of data. For
the s-channel, higher statistics are needed.  Previous CMS studies can
be found in~\cite{singletop-cms1,singletop-cms2}.


\section{Conclusions}

A summary of the potential of ATLAS and CMS in QCD and top quark
measurements has been presented, with a special emphasis on the early
data taking phase There, the focus will be on the understanding of the
detectors, and on precise measurements of known Standard Model
processes, which will pave the way for searches for new physics, for
which QCD jet and top quark events are the most important backgrounds.
CMS and ATLAS are well prepared to achieve these goals.


\end{document}